# Mathieu equation as a result of Laplace perturbation theory in the restricted three body problem

Alexey Rosaev[1] , Eva Plavalova[2]

[1]*Regional Scientific and Educational Mathematical Center Centre of Integrable System,Yaroslavl, Russia,*email: hegem@mail.ru
[2]Mathemailcal institute, Slovak academy of Science, Bratislava, Slovakia
email: plavalova@mat.savba.sk, plavalova@komplet.sk

**Abstracts**

Linear equations with periodic coefficients describe the behavior of various dynamical systems. This studying is devoted to their applications to the planetary restricted three-body problem (RTBP).
 Here we consider the Laplace method for determining perturbation in coordinates. We show that classical theory of perturbation leads to a linear equation with periodic coefficients. Than we present a modification of Laplace method. This modification allows us to study motion over a longer time interval.

### 1. Introduction

The three-body problem is a continuous source of study, since the discovery of its non-integrability by H. Poincare [1]. However, some problems still remain unresolved. In particular, the explanation of the fact that in the asteroid belt some resonances are empty while others are well populated has been until recently an open problem (Celletti et al, [2]).
 On the other hand, linear equations with periodic coefficients naturally appear in various physical and mathematical problems, so they are very common. These equations describe the behavior of various dynamical systems. This study is devoted to their application in celestial mechanics. In spite related studying are not numerous, we can note first of all Hill's lunar theory, works of Markeev [3,4]. Briefly, this problem is mentioned in Morbidelly's book [5]. The The parametric equations given by Shebeheli [6] are related to the Lindsted method. The parametric excited Hamiltonian is written in [7].

In this paper, we consider a planar restricted three-body problem, when mass m revolves around M (M is much more than *m*) in a circular orbit and the third body is considered with negligible mass $m_0$. We assume that all bodies move in the same plane.

The paper consists of four part. In the first part, the formulation of the problem, in general, is followed to the Subbotin's book [8]. In the second part of the present paper a linear equation with periodic coefficients is derived using the Laplace method of the calculation of the perturbation in coordinates. Next, a new modification of the method is proposed, which allowed to takes into account a significant part of the perturbation over a longer time interval. In the third part, some properties of the Mathieu equation solution are considered.

### 2. Problem setting and main equations

We shell start with the basic equations of motion in the restricted problem of a 3-body in a cylindrical reference frame. The cylindrical coordinate system has some advantages for many applications of planetary dynamics due to the consideration of the symmetry of the problem. The basic equations for the radius vector and longitude in the restricted problem of n-bodies in a cylindrical system are [8]:

$$\frac{d^2r}{dt^2} - r\left(\frac{d\upsilon}{dt}\right)^2 + \gamma r M \rho^{-3} = \frac{\partial U}{\partial r} \qquad (1)$$

$$\frac{d}{dt}\left(r^2 \frac{d\upsilon}{dt}\right) = \frac{\partial U}{\partial \lambda}$$

$$\frac{d^2z}{dt^2} + \gamma z M \rho^{-3} = \frac{\partial U}{\partial z}$$

$$\rho = \sqrt{r^2 + z^2}$$

Perturbation function consists of direct and indirect part:

$$U = \frac{\gamma M}{r} + \sum U_j$$

$$U_j = \frac{\gamma m}{\Delta_j} - \frac{\gamma m r}{R^2}\cos(\lambda - \lambda_j) \qquad (2)$$

$$\Delta_j = \sqrt{r^2 + R^2 - 2r R \cos(\lambda - \lambda_j)} \qquad (3)$$

where $\gamma$ – gravity constant, m – mass of perturbing body, $r(t)$ – distance small particle from central mass M, $\lambda(t), \lambda_j(t)$ – longitude of small particle and perturbing body respectively, R – radius of orbit perturbing body (Jupiter). Below we shell use Laplace method for equation linearization.

Laplace introduced a coordinate system in the plane of the osculating orbit at the moment t. He showed that, up to second-order values relative to the perturbing forces, the longitude can be replaced by the longitude $\lambda$ in the fixed plane corresponding to the moment $t_0$.

After that, equations (1) can be reduced to form:

$$\frac{d^2r}{dt^2} - r\left(\frac{d\lambda}{dt}\right)^2 + \gamma M r^{-2} = \frac{\partial U}{\partial r} \qquad (4)$$

$$\frac{d}{dt}\left(r^2 \frac{d\lambda}{dt}\right) = \frac{\partial U}{\partial \lambda}$$

### 3. Laplace method for calculation of perturbations in coordinates

#### 3.1. The Classical Laplace Method

In this section we will describe the Laplace method following by Subbotin book [8]. The classical Laplace's method for calculating perturbations in coordinates is aimed to describe motion in a short time interval, mainly along one revolution of a body along its orbit. As a result, Laplace reduced the problem of calculating perturbations in coordinates to the construction of a perturbation theory for (independent) equations (see [8] for details):

$$\ddot{q}_i + \gamma M r_{i-1}^{-3} q_i = Q_i \qquad (6)$$

where $q_i = (r_{i-1}\delta r)$, $r_i = r_i(t) = r_{i-1} + \delta r_{i-1}(t)$  $Q = Q(t) = r_0 \dfrac{\partial U}{\partial r_0} + 2U + C$.

After the solution of the equation (6) for i-approach, the resulted values of $r_i$ substituted in Q and calculations repeated. After the substitution the first order approximation:

$$\delta r = C_1 \cos(\omega t + f)$$
$$r_1^{-3} = (r_0)^{-3} - 3\cos(\omega t + f)(r_0)^{-4} C_1 + ..., \qquad (7)$$

in the equation (6) we have:

$$\ddot{q} + \gamma M (r_0)^{-3}(1 - 3\cos(\omega t + f) C_1 / r_0) q = Q \qquad (8)$$

The equation (8) may be rewritten in a form:

$$\ddot{q} + \omega_0^2 (1 + h \cos \omega t) q = Q(t) \qquad (9)$$

where:

$$\omega_0^2 = \gamma M r_0^{-3} \qquad (10)$$

$$h = -3 C_1 / r_0 \qquad (11)$$

Already the second approximation gives Mathieu equation. The equation for $\delta \lambda$ determination has a form [8]:

$$2 r_0^2 \frac{d\lambda_0}{dt} \frac{d\delta\lambda}{dt} + \delta r \left( \frac{d^2 r_0}{dt^2} \right) - r_0 \frac{d^2 \delta r}{dt^2} + \frac{3\gamma M r_0 \delta r}{r_0^3} = -r_0 \frac{\partial U}{\partial r} \bigg|_{r=r_0} + ... \qquad (12)$$

The classic Laplace method calculation of the perturbation in coordinates aimed to describe motion on the short time interval mostly along one revolution body in its orbit. On each step the partial solution of non-homogenous equation (6) is necessary to looking for, which it not suitable. As a result, this method cannot give the proper picture of evolution in the long time interval. By this reason we made some modification of Laplace method.

### 3.2. The modified Laplace method

Come back to equation (6). Note, that perturbation function possible to expand by power $\delta r \equiv \dfrac{q}{r_0} \ll 1$. By using:

$$Q(t) = \frac{\partial^2 U}{\partial r_{i-1}^2} q_i + 2 \frac{\partial U}{\partial r_{i-1}} q_i / r_{i-1} + C_0 + ... \qquad (13)$$

Here $q_i = (r_{i-1}\delta r)$. The equation (6) becomes:

$$\ddot{q}_i + \omega^2 q_i = C_0 \tag{14}$$

In the first approach the equation (14) becomes:

$$\ddot{q} + \gamma M(r_0)^{-3}(1 - 3\cos(vt+f)C_1/r_0)q - \frac{\partial^2 U}{\partial r_0^2}q - \frac{2}{r_0}\frac{\partial U}{\partial r_0}q = Q' \tag{15}$$

it is equation (9) where:

$$\omega_0^2 = \gamma M(r_0)^{-3}\left(1 - \frac{\partial^2 U}{\partial r_0^2} - \frac{2}{r_0}\frac{\partial U}{\partial r_0}\right) \tag{16}$$

$$h = 3\gamma M C_1(r_0)^{-4}\left(1 - \frac{\partial^2 U}{\partial r_0^2} - \frac{2}{r_0}\frac{\partial U}{\partial r_0}\right)^{-1} \tag{17}$$

This modification makes it possible to qualitatively study evolution over a time interval longer than in the classical method.

### 4. The Mathieu equation solution

Due to the great importance of Mathieu equations in applications, we will say a few words about their solution. The solution is not difficult to obtain numerically.

As it is well known, there are two types of solution of the equation (9): a stable close-to-periodic solution (Fig.1) and a solution with unbounded amplitude (Fig.2). Therefore, at some initial conditions, the instability of the orbits can appear. Also, a modification of stable solution with two frequencies is known (Fig.3). The last figure is very similar to the evolution of the orbital elements of real small bodies close to resonance (see, for example, [9]).

The condition of the parametric resonance becomes (Landau & Lifshitz, [10]):

$$\omega = 2\omega_0/K = jl/(k+l)n \tag{18}$$

where n is mean motion, j, l, k, K are integer. The parametric resonance has notable width only when K=1 or K=2. The width of instable area is determined by the coefficient h (Landau & Lifshitz, [10]):

$$-h/2\omega_0 < \delta < h/2_j\omega_0 \tag{19}$$

Previously [12], we note that orbits near resonances (2n+1)/(2n-1) are instable parametric. This conclusion completely coincides with results of the studies of the declared problem in the paper Hadjidemetriou [11].

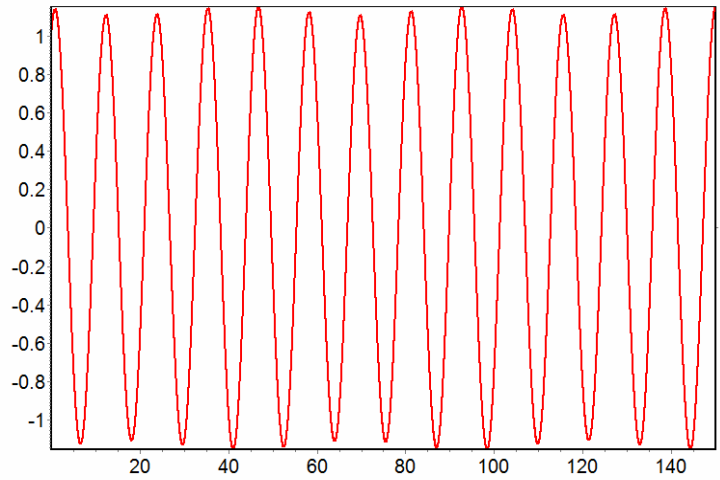

Fig.1

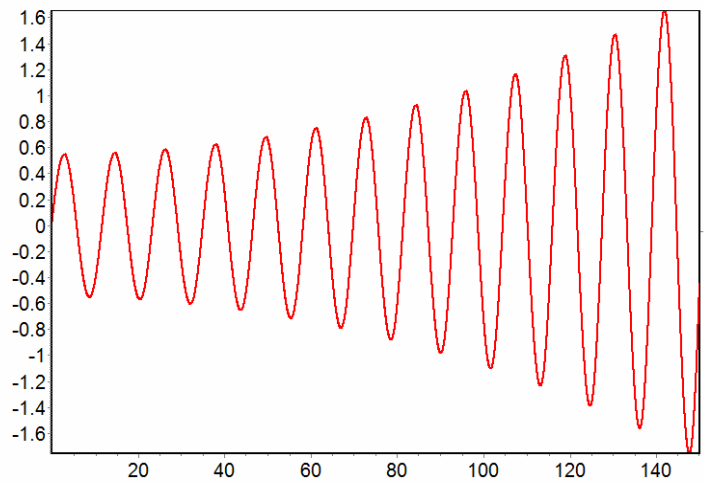

Fig.2

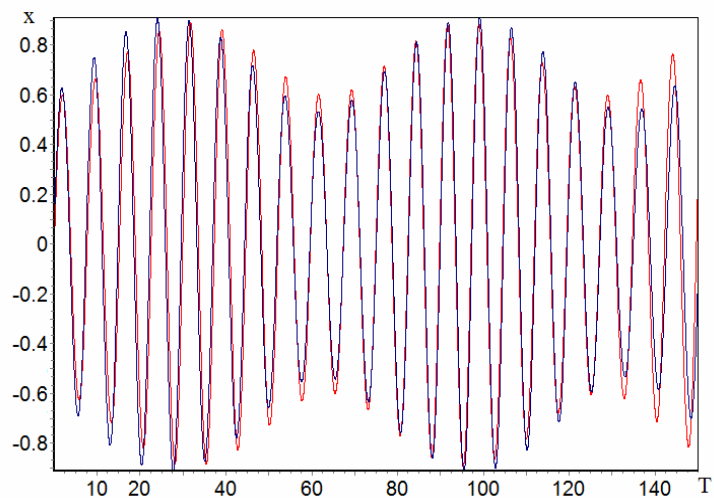

Fig.3

However, the position of the instable zone can depend on eccentricity (Rosaev, [12]). If confirmed, the dependence can help to explain the stability of some resonant asteroids with large eccentricity (Fig.4). Also it can be useful in study the process of resonance overlap. However, the new numerical and theoretical study is required.

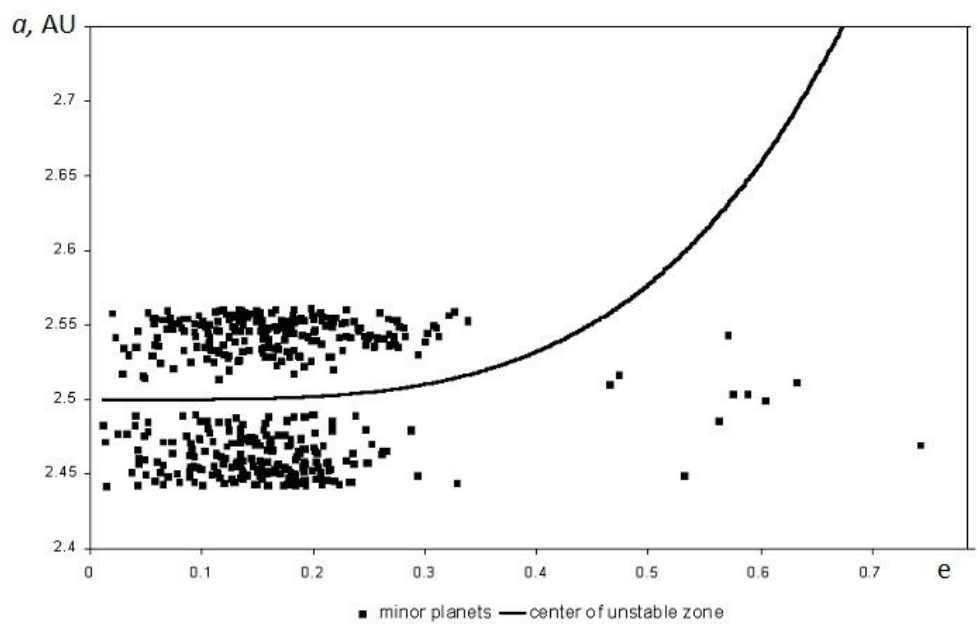

Fig.4. The distribution minor planets close 3:1 resonance

Conclusions

Linear equations with periodic coefficients describe the behavior different dynamical systems. This studying is devoted to their applications in celestial mechanics.

We consider the Laplace method of perturbation in coordinates. We show that classical theory of perturbation leads to the linear equation with periodic coefficients. Here we give the modification of Laplace method of perturbation in coordinates. This modification allows us to study motion over a longer time interval.

In some practically interesting cases, we can reduce problem to Mathieu equations. Due to the great importance of Mathieu equations in applications, we discuss the character of its solution and the conditions of stability.